\documentclass[fleqn,12pt,twoside]{article}
\usepackage{espcrc1}

\usepackage{graphicx}
\usepackage[figuresright]{rotating}

\newcommand{\AmS}{{\protect\the\textfont2
  A\kern-.1667em\lower.5ex\hbox{M}\kern-.125emS}}

\title{Heavy Ion Physics with CMS at LHC}

\author{Pablo P. Yepes
              \address{Physics and Astronomy Department\\
	      MS 315, Rice University \\
	      Houston, TX 77005, USA\\
	      http://yepes.rice.edu\\
	      yepes@rice.edu}
	      \\ for the CMS Collaboration
	      }
\begin{document}

\maketitle

\begin{abstract}
The CMS detector is well equipped to provide unique 
measurements in heavy ion
collisions at LHC. It will provide measurements of the J/$\Psi$ and $\Upsilon$
families with good separation of the different resonances. Jet 
quenching
will be studied by analyzing monojet-to-dijet ratios and  $Z^0$($\gamma$)+jet
events.
\end{abstract}
\section{CMS Detector}
The Large Hadron Collider (LHC) plans to operate as a heavy ion collider for
one month each year starting in 2007. LHC will be capable of accelerating a
variety of ions up to beam energies of 7 TeV/charge. The design, peak 
and three-hour-fill-averaged
luminosities are presented in Table 1 along with the maximum center of mass 
energy\cite{brandt}. 
\begin{table}[hbt]
\begin{center}
\begin{tabular}{|c|c|c|c|}
\hline
                          & $L_{max}$        & $<$L$>$            & $\sqrt{s_{NN}}$ \\
 Ion                      & $(cm^{-2}s^{-1})$& $(cm^{-2}s^{-1})$  & (GeV)    \\
\hline
$^{208}Pb^{82}$           & 1.0 $10^{27}$  & 4.2 $10^{26}$ & 5500 \\
$^{120}Sn^{50}$           & 1.7 $10^{28}$  & 7.6 $10^{27}$ & 5800 \\
$^{84}Kr^{36}$            & 6.6 $10^{28}$  & 3.2 $10^{28}$ & 6000 \\
$^{40}Ar^{18}$            & 1.0 $10^{30}$  & 5.2 $10^{29}$ & 6300 \\
$^{16}O^{8}$              & 3.1 $10^{31}$  & 1.4 $10^{31}$ & 7000 \\
\hline
\end{tabular}
\end{center}
\caption{LHC planned ion beams with design, maximum and fill-averaged luminosities along with
maximum center of mass energies.}
\label{lhcParameters}
\end{table}
The Compact Muon Solenoid (CMS) \cite{cms} experiment is a general-purpose facility to
study hadronic collisions at the LHC. The detector
consists of a tracking system, electromagnetic and hadronic calorimeters, and muon detectors. A
solenoidal magnet providing a 4 Tesla magnetic field surrounds the tracking and calorimetric
systems. The
tracker, covering the rapidity region $|{\eta}|<$ 2.5, is based on silicon technology. The
electromagnetic calorimeter consists of about 83000 lead-tungstate crystals arranged in 
a central barrel covering
$|{\eta}|<$1.48 and the endcaps, which extend its range to rapidity $|{\eta}|<$3. In the central
barrel, the granularity is as high as ${\eta}$x$\phi$ = 0.0175x0.0175. The hadronic calorimeter consists
of barrel and endcap sections, each made of sandwiches of copper plates and plastic
scintillator. In the central region ($|{\eta}|<2$) the ${\eta}$x$\phi$ segmentation is 0.087x0.087. The
combination of electromagnetic and hadronic calorimeters provides coverage of the central
rapidity region with excellent energy resolution. In addition, coverage at large rapidity
($3<|\eta|<5$) is achieved by two very forward calorimeters. The large calorimetric coverage 
and good energy resolution makes CMS an optimal detector for jet studies. 

The muon system covers the $|\eta|<2.5$ region. 
In the barrel ($|{\eta}|<1.5$), muons must have a
transverse momentum larger than 3.5 GeV/$c^2$ to be efficiently detected.  

\section{Quarkonia Production}
The suppression of charmonium production has provided strong evidence for the
formation of dense hot hadronic matter at SPS energies\cite{charmonium}. 
At RHIC,
$J/\Psi$ production is an important part of the PHENIX\cite{zajc} physics program.
STAR is also expected to study charmonium production\cite{LeCompte:1999sf}. At LHC, CMS
is well equipped to study bottonium and charmonium production.
The ability to study $\Upsilon$ yields at LHC,
which has a negligible cross sections at lower energies, will allow
further understanding of the characteristics of dense hadronic matter.

\begin{figure}[htb]
\begin{center}
\includegraphics[scale=0.50]{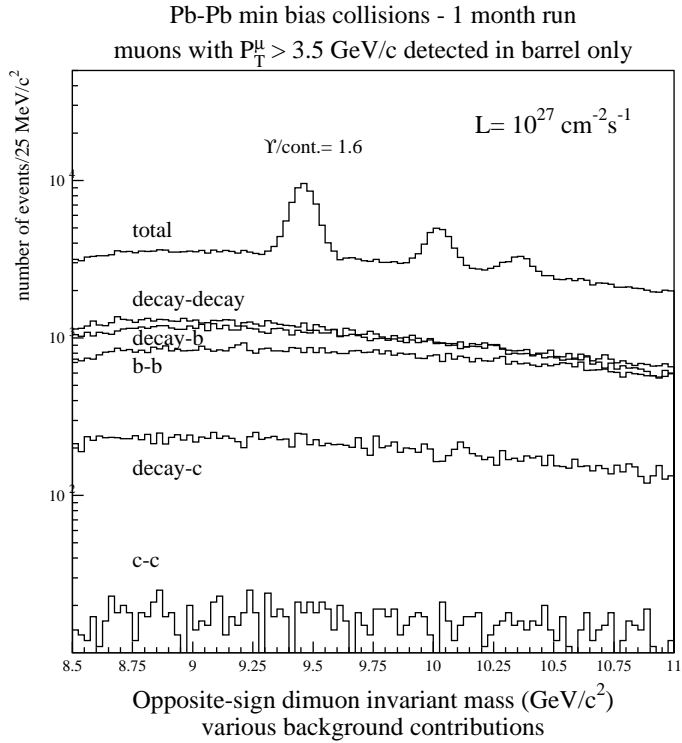}
\end{center}
\caption{Invariant dimuon spectrum reconstructed with the barrel muon system 
expected in one month of running with PbPb. The background from different 
sources is indicated along with the sum of all contributions.}
\label{upsilon_pb}
\end{figure}

A detailed study of the feasibility of quarkonia detection in CMS was
carried out. Production cross sections were estimated utilizing the 
expression: $\sigma_{AA} = A^{2\alpha} \sigma_{pp} $ with $\alpha$=0.9 (0.95) for
$J/\Psi$($\Upsilon$). $\sigma_{pp}$ was obtained by extrapolating Tevatron
data from $\sqrt{s}$=1.8 to 5.5(7) TeV for PbPb(CaCa). The obtained cross sections
are shown in Table \ref{tableCrossSections}.
\begin{table}[hbt]
\begin{center}
\begin{tabular}{|c|r|r|r|r|r|r|}
\hline
&\multicolumn{2}{c}{Cross Sections}&\multicolumn{4}{|c|}{Reconstructed}\\ 
&&&\multicolumn{4}{|c|}{in 1 Month Run}\\ \hline
Ion & PbPb & CaCa &\multicolumn{2}{c}{PbPb      }&\multicolumn{2}{|c|}{CaCa}\\ \hline
&(mbarns)&($\mu$barns)&\# Events & S/$\sqrt{B}$ &\# Events& S/$\sqrt{B}$\\ \hline
$J/\Psi$        & 58.  & 3600   & 10600 & 100 & 220000 & 1440 \\
$\Psi'$         & 1.4  & 90     &   350 & 2.4 &   5800 &   92 \\
$\Upsilon$      & 0.4  & 21     & 22000 & 188 & 340000 & 1788 \\
$\Upsilon'$     & 0.12 & 6.4    &  7500 &  64 & 115000 &  605 \\
$\Upsilon''$    & 0.04 & 2.1    &  2500 &  21 &  37734 &  198 \\
\hline
\end{tabular}
\end{center}
\caption{Quarkonia cross sections at LHC along with the projected number of reconstructed
events in CMS in the dimuon channel and the signal over the square root of the background.}
\label{tableCrossSections}
\end{table}
The muon system acceptance for quarkonia decaying into dimuons  was 
studied using the barrel and the endcap detectors. In the barrel,
only $J/\Psi$ particles with $p_T>5$ GeV/c are detected, with an acceptance
at the level of a few per cent. When the endcap chambers are included,
$J/\Psi$ with $p_T=0$
is detected with an acceptance of 2\%. For $\Upsilon$, the acceptance
is 30(15)\% when the end-cap detectors are (are not) considered, with little
dependence on $p_T$. Quarkonia reconstruction was studied with detailed
simulation and full reconstruction programs. 
Muon background from $\pi$, K and heavy quark leptonic decays was considered.
The $p_T$ spectra for pions and kaons was
obtained from SHAKER \cite{shaker}. The average $p_T$, 0.48(0.67) GeV/c for
pion(kaons), is higher than that obtained with other event generators. 
The multiplicity in central PbPb events is assumed to be
$dN/dy=8000$. 
Given that
our estimates are pessimistic, backgrounds are probably overestimated.
Fig.~\ref{upsilon_pb} depicts the expected  dimuon invariant mass spectrum
in the $\Upsilon$ region
for 1 month of PbPb running. 
Table \ref{tableCrossSections} shows the number of expected reconstructed events
for the J/$\Psi$ and $\Upsilon$ families with PbPb and with CaCa in
one month running at design luminosity. In addition,
the statistical significance of the signal, number of particles divided by the 
square root of the background, is also given. 
\section{Jet Studies}
Recent results from RHIC \cite{harris,zajc} show a softening of the transverse momentum
above 5 GeV/c in central collisions. This could be attributed to jet quenching in
the hot hadronic media produced after the collision. 
At LHC, $10^7$ dijets with $E^{jet}_T>$100 GeV are produced in the region $|\eta|<$2.6 
in a one-month run of PbPb. The number is reduced by about a factor of 2 if only the
barrel is considered, $|\eta|<$1.5. 
At LHC, jet reconstruction in PbPb is possible, even for central collisions.
A modified version of the UA1 jet-finding algorithm was used to reconstruct jets in 
central PbPb collisions using the information provided by the electromagnetic and
hadronic calorimeters. As in the quarkonia study, 
the multiplicity in central PbPb events is assumed to be
$dN/dy=8000$. 
\begin{table}[hbt]
\begin{center}
\begin{tabular}{|c|c|c|c|}
\hline
 $E_T^{min}$ (GeV)        & Efficiency(\%)   & Contamination (\%) & $\sigma(E_T)/E_T$(\%) \\
\hline
50                        & 94$\pm$3          & 12.0$\pm$3    & 16.7 \\
100                       &100$\pm$2          &  1.0$\pm$0.4  & 11.6 \\
150                       & 98$\pm$2          &  0.4$\pm$0.3  &  9.2 \\
200                       & 99$\pm$2          &  0.4$\pm$0.3  &  8.6 \\
\hline
\end{tabular}
\end{center}
\caption{Efficiency, contamination, and energy resolution for jets above different transverse 
momentum thresholds in central PbPb collisions.}
\label{jetReco}
\end{table}
Table \ref{jetReco} shows the efficiency, contamination and energy resolution for finding
jets above certain thresholds. As can be seen, even the lowest threshold jets
can be reconstructed with good efficiency and relatively low background.
This good performance will allow CMS to study 
jet quenching in dijet production \cite{gyulassy}.
Jet energy loss in hadronic matter can also be measured by
tagging jets opposite to non-strongly interacting particles such as $Z^0$ or gamma,
both produced in the reactions $qg\rightarrow qV$ and $q\bar{q}\rightarrow gV$, where
V=$Z^0$ or $\gamma$.
\section{Conclusions} 
In conclusion, CMS provides a unique tool to study heavy ion collisions at the LHC.
It will probe hot hadronic matter by studying J/$\Psi$ and $\Upsilon$ production.
Its excellent calorimetry will provide
large coverage and good energy resolution for jet quenching studies.


\begin{thebibliography}{9}
\bibitem{brandt} D.~Brandt, {\it $5^{th}$ CMS Heavy Ion Meeting}, St. Petersburg, June 11-14, 
2000.
\bibitem{cms} The Compact Muon Solenoid, Technical Proposal, CERN/LHCC 94-38, LHCC/P1, 
Dec 15, 1994.
\bibitem{charmonium}
M.~C.~Abreu {\it et al.}  [NA50 Collaboration],
Phys.\ Lett.\ B {\bf 477}, 28 (2000).
\bibitem{zajc} W.~Zajc, Recent Results from PHENIX, in these proceedings. 
\bibitem{LeCompte:1999sf} T.~J.~LeCompte  [STAR Collaboration],
{\it Workshop on Quarkonium Production in Relativistic Nuclear Collisions}, Seattle, WA,
11-15 May 1998..
\bibitem{shaker} F.~Antinori, CERN/ALICE 93-01 (MC).
\bibitem{harris} J.~Harris, Recent Results from STAR, in these proceedings.
\bibitem{gyulassy} M. Gyulassy and M. Plumer, Phys.\ Lett.\ B {\bf 243}, 432 (1990).
                 
\end{thebibliography}
\end{document}